\documentclass[fleqn]{article}
\usepackage[centertags]{amsmath}
\usepackage{espcrc2}
\usepackage{graphicx}
\title{Indications on the Higgs boson mass from lattice
simulations }

\author{P. Cea\address[DF,INFN]{Dipartimento di Fisica, Univ. of Bari and INFN - Sezione di Bari,
        I-70126 Bari, Italy},
        M. Consoli\address[INFNC]{INFN - Sezione di Catania, I-70126 Catania, Italy}  and
        L. Cosmai\address[INFN]{INFN - Sezione di Bari, I-70126 Bari, Italy}}

\begin{document}

\begin{abstract}
The `triviality' of $\Phi^4_4$ has been traditionally interpreted
within perturbation theory where the prediction for the Higgs
boson mass depends on the magnitude of the ultraviolet cutoff
$\Lambda$. This  approach crucially assumes that the vacuum field
and its quantum fluctuations rescale in the same way. The results
of the present lattice simulation, confirming previous numerical
indications, show that this assumption is not true. As a
consequence, large values of the Higgs mass $m_H$ can coexist with
the limit $\Lambda\to \infty $. As an example, by extrapolating to
the Standard Model our results obtained in the Ising limit of the
one-component theory, one can obtain a value as large as $m_H=760
\pm 21$ GeV, independently of $\Lambda$. \vspace{1pc}
\end{abstract}

\maketitle
Traditionally, the `triviality' of $\Phi^4$ theories in $3+1$
space-time dimensions~\cite{Sokal_book} has been interpreted
within perturbation theory. In this {\it interpretation}, these
theories represent just an effective description, valid only up to
some cutoff scale $\Lambda$.
This conventional view, when used in the Standard Model, leads to
predict that the Higgs boson mass squared, $m^2_H$, is
proportional to $g_R v^2_R$, where $v_R$ is the known weak scale
(246~GeV) and $g_R \sim 1/{\ln \Lambda}$ is the renormalized
scalar self-coupling. Therefore, the ratio $m_H/v_R$ would be a
cutoff-dependent quantity that becomes smaller and smaller when
$\Lambda$ is made larger and larger. However,  in an alternative
approach~\cite{Consoli:1994jr,Consoli:1999ni} this conclusion is
not true. The `Higgs condensate' and its quantum fluctuations
undergo different rescalings when changing the ultraviolet cutoff,
so that the relation between $m_H$ and the physical $v_R$ is not
the same as in perturbation theory.

To understand this point, we observe that beyond perturbation theory, in
a broken-symmetry phase, there are two different definitions of
the field rescaling. There is a rescaling of the `condensate', say
$Z\equiv Z_\varphi$, and a rescaling of the fluctuations, say
$Z\equiv Z_{\text{prop}}$.
Consider a one-component scalar theory and introduce the bare
expectation value $v_B=\langle\Phi_{\text{ latt}}\rangle$
associated with the `lattice' field as defined at the cutoff
scale. By $Z\equiv Z_\varphi$ we mean the rescaling that is needed
to obtain the physical vacuum field $v_R= v_B / \sqrt{Z_\varphi}$.
Since the second derivative of the effective potential is the
zero-four-momentum two-point function, this standard definition is
equivalent to define $Z_\varphi$ as:
\begin{equation}
\label{z1phi} Z_\varphi= m^2_H \chi_2(0)
\end{equation}
where $\chi_2(0)$ is the zero-momentum susceptibility.
On the other hand, $Z\equiv Z_{\text{prop}}$ is determined from
the residue of the connected propagator on its mass shell.
Assuming `triviality' and the K\'allen-Lehmann representation for
the shifted quantum field, one predicts $Z_{\text{prop}} \to 1$
when approaching the continuum theory.

Now, in the standard approach
one  assumes  $Z_\varphi\sim Z_{\text {prop}}$ while in
the different interpretation of triviality,
\cite{Consoli:1994jr,Consoli:1999ni}
although $Z_{\text
{prop}}\to 1$, as in leading-order perturbation theory,
$Z_\varphi\sim \ln \Lambda $ is fully non perturbative and
diverges in the continuum limit. In this case,
in order to obtain $v_R$ from the
bare $v_B$ one has to apply a non-trivial correction. As a result,
$m_H$ and $v_R$ now scale uniformly in the continuum limit, and
the ratio $C=m_H/v_R$ is a cutoff-independent quantity.
To check this alternative picture against the generally accepted
point of view, one can run numerical simulations of the theory. In
this respect, we observe that numerical evidence for different
cutoff dependencies of $Z_\varphi$ and $ Z_{\text{prop}}$ has
already been already
reported~\cite{Cea:1998hy,Cea:1999kn,Cea:1999zu}. In those
calculations,
one was fitting the lattice data for the connected
propagator to the (lattice version of the) two-parameter form
\begin{equation}
\label{gprop} G_{\text{fit}}(p)= \frac{Z_{\text{prop}}}{ p^2 +
m^2_{\text{latt}} }.
\end{equation}
After computing the zero-momentum susceptibility
$\chi_{\text{latt}}$, it was possible to compare the value of
$Z_\varphi \equiv m^2_{\text{latt}} \chi_{\text{latt}}$
 with the fitted $Z_{\text{prop}}$,
both in the symmetric and broken phases. While no difference was
found in the symmetric phase, $Z_\varphi$  and $Z_{\text{prop}}$
were found to be sizeably different in the broken phase.
$Z_{\text{prop}}$ was very slowly varying and steadily approaching
unity from below in the continuum limit. On the other hand,
$Z_{\varphi}$ was found to rapidly increase {\it above} unity in
the same limit.
A possible objection to this strategy is that the two-parameter
form  Eq.~(\ref{gprop}), although providing a good description of
the lattice data, neglects higher-order corrections to the
structure of the propagator. As a consequence, one might object
that the extraction of the various parameters is affected in an
uncontrolled way. For this reason, we have decided to change
strategy by performing a new set of lattice calculations. Rather
than studying the propagator, we have addressed the
model-independent lattice measurement of the susceptibility. In
this way, {\it assuming} the mass values from perturbation theory,
one can obtain a precise determination of $Z_\varphi$ to be
compared with the perturbative predictions.

The numerical simulations  were performed in the Ising limit
where a one-component $(\lambda\Phi^4)_4$ theory becomes
\begin{equation}
\label{ising}
S_{\text{Ising}} = -\kappa \sum_x\sum_{\mu} \left[ \phi(x+\hat
e_{\mu})\phi(x) + \phi(x-\hat e_{\mu})\phi(x) \right]
\end{equation}
and $\phi(x)$ takes only the values $\pm 1$ (in an infinite
lattice, the broken phase is found for $\kappa > 0.07475$).
Using the Swendsen-Wang  and
 Wolff cluster algorithms we have computed
%
%
%
the zero-momentum susceptibility:
\begin{equation}
\label{chi}
 \chi_{\text{latt}}=L^4 \left[ \left\langle |\phi|^2
\right\rangle - \left\langle |\phi| \right\rangle^2 \right] .
\end{equation}
We used different lattice sizes at each value of $\kappa$ to have
a check of the finite-size effects. Statistical errors have been
estimated using the jackknife. Finally, we have checked our
results with those obtained by other authors~\cite{Jansen:1989cw}.
%
%
%
%

%
%
As anticipated, we shall use the perturbative predictions for the
Higgs boson mass adopting the
L\"uscher-Weisz scheme~\cite{Luscher:1988ek}.
To this end, let us denote by
$m_{\text{input}}$ the value of the parameter $m_R$ reported in
the first column of Table~3 in Ref.~\cite{Luscher:1988ek} for any
value of $\kappa$ (the { Ising limit corresponding to the value of
the other parameter $\bar{\lambda}=1$). In this way, one can compare the
quantity
\begin{equation}
\label{zphi}
       Z_\varphi\equiv 2\kappa m^2_{\text{input}} \chi_{\text{latt}}
\end{equation}
with the perturbative prediction for $Z_{\text{LW}}\equiv 2\kappa
Z_R$  where $Z_R$ is defined in the third column of Table~3 in
Ref.~\cite{Luscher:1988ek}.

\begin{figure}[t]
\begin{center}
\includegraphics[width=8.0cm,clip]{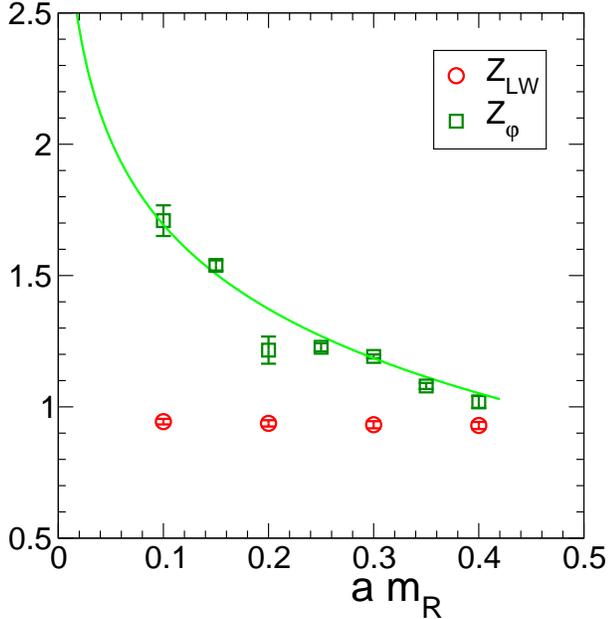}
\vspace{-1.1cm} \caption{The values of $Z_\varphi$ as defined
through Eq.~(\ref{zphi}) and $Z_{\text{LW}}$ versus
$m_{\text{input}}=a m_R$. Solid line is the fit of $Z_\varphi$
according to $Z_\varphi = \text{constant}+B \ln{\frac{1}{am_R}}$.}
\end{center}
\end{figure}
The values of $Z_\varphi$ and
$Z_{\text{LW}}$ for various $\kappa$ are reported in Fig.1.
As one can check, the two $Z$'s follow completely different trends
and the discrepancy becomes larger and
larger when approaching the continuum limit, precisely the same
trend found in Refs.\cite{Cea:1999kn,Cea:1999zu}. This confirms
that, approaching the continuum limit, the rescaling of the `Higgs
condensate' cannot be described in perturbation theory.
In addition, the lattice data
for $Z_\varphi$
are completely consistent with the alternative scenario
$\sim \ln \Lambda$ predicted in
Refs.\cite{Consoli:1994jr,Consoli:1999ni}.

Now, if the physical $v_R$ has to be computed from the bare $v_B$
 through $Z=Z_\varphi$, rather than
through the perturbative $Z=Z_{\text{LW}}$, one may wonder about the
$m_H$-$v_R$ correlation. In this case the perturbative relation
\begin{equation}
\label{gR} \left[ \frac{m_H}{v_R} \right]_{\text{LW}} \equiv \sqrt
{  \frac{g_R}{3} }.
\end{equation}
becomes
\begin{equation}
\label{mh}
\frac{m_H}{v_R}= \sqrt{ \frac{g_R}{3}
\frac{Z_\varphi}{Z_{\text{LW}} } } \equiv C
\end{equation}
obtained by replacing $Z_{\text{LW}} \to Z_\varphi$ in
Ref.~\cite{Luscher:1988ek} but correcting for the perturbative
$Z_{\text{LW}}$  introduced in the L\"uscher and Weisz approach.
%
%
%
%
Assuming the values of
$g_R$ reported in the second column of
Table~3 of Ref.~\cite{Luscher:1988ek} and using our values of $Z_\varphi$,
we have reported in Fig.2
the values of $m_H$ as defined through
 Eq.~(\ref{mh}) versus $m_{\text{input}}=a m_R$ for $v_R=246$ GeV.
The error band corresponds to a one standard deviation
error in the determination of $m_H$ through a fit with a constant
function. As one can see, the $Z_\varphi \sim \ln \Lambda$ trend observed
in Fig.1,
compensates the $1/\ln \Lambda$ from $g_R$ so that
$C$ turns out to be a
cutoff-independent constant.

%
%
%
%
%
%
%
%
%
\begin{figure}[t]
\begin{center}
\includegraphics[width=8.0cm,clip]{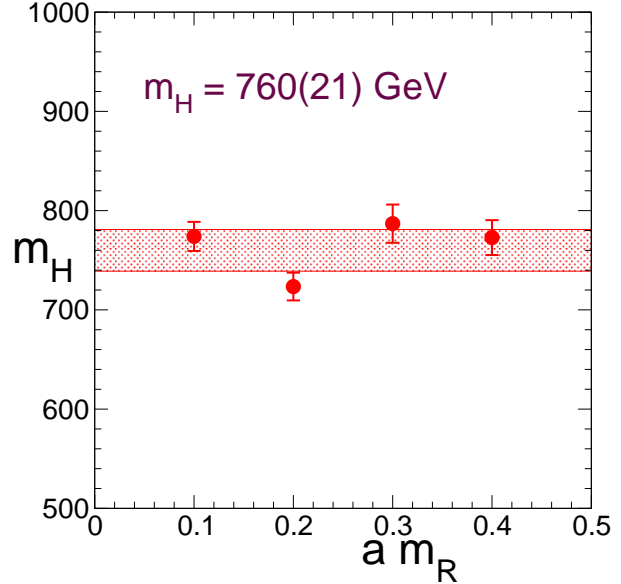}
\vspace{-1.1cm} \caption{$m_H$ as defined through
 Eq.~(\ref{mh}) versus $m_{\text{input}}=a m_R$}
\end{center}
\end{figure}
%
%
%
%
Our results imply that the value of the  Higgs boson mass, in
units of $246$ GeV, {\it does not depend} on the magnitude of the
ultraviolet cutoff. Therefore,  the whole issue of the upper
bounds on the Higgs mass is affected suggesting the need of more
extensive studies of the critical line to compare the possible
values of $ C $ in the full 2-parameter $ \Phi^4_4 $ theory.

In any case, a value as large as $ m_H=760 \pm 21 $~GeV, would
also be in good agreement with a recent phenomenological analysis
of radiative corrections~\cite{Loinaz:2002ep} that points toward
substantially larger Higgs masses than previously obtained through
global fits to Standard Model observables.


\end{document}